\documentclass{article}
\newcommand{\undertitle}{}
\newcommand{\shorttitle}{\textbf{LogDoctor}}
\usepackage{arxiv}
\usepackage{float}
\title{LogDoctor: an open and decentralized worker-centered solution for occupational management in healthcare}
\date{\today}
\author{ \href{https://orcid.org/0000-0002-9073-4504}{\includegraphics[scale=0.06]{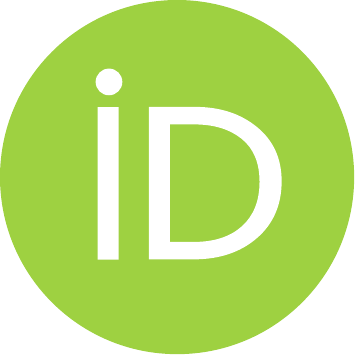}\hspace{1mm}Sami Barrit}\\
	Neurochirurgie, Université Libre de Bruxelles, Belgium\\
        Sciences Chirurgicales, Université Paris-Est Créteil, France\\
	Sciense, New York, United States\\
	\texttt{sami@sciense.org} \\
	\And
	Alexandre Niset \\
	Médecine d'Urgence, Université Catholique de Louvain, Belgium\\
	Sciense, New York, United States \\
}

\usepackage{threeparttable}

\usepackage{fancyhdr}
\usepackage[hidelinks]{hyperref}
\usepackage[english]{babel}

\usepackage{hyperref,amssymb,amsmath,graphicx,enumitem,listings,verbatim,eurosym, fancyhdr, array}

\usepackage[utf8]{inputenc}
\usepackage{tikz,tikz-qtree, tikz-cd}
\usepackage{booktabs, makecell, multirow,multicol}
\usepackage{xcolor}
\usepackage{colortbl}
\usepackage{tabularx}
\usepackage{etoolbox}
\usepackage{colortbl}

\newcolumntype{P}{>{\centering}p{2.5cm}}

\usetikzlibrary{shapes, arrows, calc, arrows.meta, fit, positioning,shapes.symbols,decorations.pathreplacing, arrows.meta, patterns,shadows.blur, mindmap, trees}

\definecolor{asparagus}{rgb}{0.53, 0.66, 0.42}
\definecolor{bleudefrance}{rgb}{0.19, 0.55, 0.91}
\definecolor{orange}{rgb}{0.98, 0.6, 0.01}
\definecolor{bostonuniversityred}{rgb}{0.8, 0.0, 0.0}
\definecolor{darkpastelpurple}{rgb}{0.59, 0.44, 0.84}
\definecolor{darkpastelgreen}{rgb}{0.01, 0.75, 0.24}

\definecolor{Tundora}{rgb}{0.29,0.29,0.29}
\definecolor{Silver}{rgb}{0.749,0.749,0.749}
\DeclareUnicodeCharacter{0301}{*************************************}

\usepackage{setspace}
\usepackage{csquotes}
\usepackage[sorting=none, backend=biber]{biblatex}
\begin{document}

\maketitle
\begin{abstract} Occupational stress among health workers is a pervasive issue that affects individual well-being, patient care quality, and healthcare systems' sustainability. Current time-tracking solutions are mostly employer-driven, neglecting the unique requirements of health workers. In turn, we propose an open and decentralized worker-centered solution that leverages machine intelligence for occupational health and safety monitoring. Its robust technological stack, including blockchain technology and machine learning, ensures compliance with legal frameworks for data protection and working time regulations, while a decentralized autonomous organization bolsters distributed governance. To tackle implementation challenges, we employ a scalable, interoperable, and modular architecture while engaging diverse stakeholders through open beta testing and pilot programs. By bridging an unaddressed technological gap in healthcare, this approach offers a unique opportunity to incentivize user adoption and align stakeholders' interests. We aim to empower health workers to take control of their time, valorize their work, and safeguard their health while enhancing the care of their patients.
\end{abstract}
\section{Background}

Work-related stress is a long-standing and widespread public health issue \cite{leka2003work}, associated with inadequate working conditions \cite{murphy1995occupational} due to structural failures \cite{pross2010culture}. Health workers are particularly exposed due to common resource-constrained \cite{Sauter1999} and crisis settings \cite{prasad2021prevalence}, as well as the inherent physical and emotional burdens of caregiving \cite{Ruotsalainen2014}. Primary missions such as caring, education, and research are often impeded by cost-cutting measures reducing supportive resources and the increasing administrative burden from medical liability documentation. These mismatches between resources and capacities result in work overload \cite{nirel2008stress} \cite{skinner2008work} \cite{michie2002causes}, job dissatisfaction \cite{Aiken2002}, and work-life imbalance \cite{humphries2020hospital}.

On a personal level, this leads to loss of health and life due to burnout, depression, suicide, stroke, and cardiovascular diseases \cite{greiner2004occupational} \cite{huerta2013effects} \cite{ferguson1973study} \cite{Steptoe2013}. On a collective level, it compromises the population's safety by undermining the quality of care \cite{moustaka2010sources} and contributes to the brain drain of healthcare professionals and staff shortages, creating a self-reinforcing cycle \cite{humphries2020hospital}. This growing issue threatens both the human and the economic sustainability of healthcare systems \cite{montgomery_lainidi_2023} \cite{pawelczyk2018stress}.

Various institutional, national, and supranational initiatives have been undertaken to tackle this challenge \cite{burchiel20172017} \cite{sundberg2014swedish}. However, even legal regulations have shown their limits in changing poor working conditions in healthcare \cite{Clarkee004390} \cite{cresswell2009optimising} \cite{tait2008current} \cite{machtey2018regulations}, partly due to the shortcomings of most working time recording systems. These employer-driven solutions focus on shift planning, rostering, and billing but are not designed to ease time tracking for health workers or ensure accurate activity tracking and safety monitoring \cite{StaffSite}   \cite{PetalSite} \cite{TribeSite}. Additionally, the recordings may be subject to opaque management and manipulation, further exacerbating the issue, as they are particularly exposed to peer, hierarchical, and socio-cultural pressures \cite{hicss-2016}.

Here, we propose an open and decentralized worker-centered solution utilizing machine intelligence. Our objective is to provide an empowering solution for health workers to take control of their time, valorize their work, enhance their capacity, and safeguard their health and that of their patients despite socio-cultural or economic pressures. 

\section{Conceptual framework}

The input consists of multimodal data provided by the health worker, employer, or an approved third party. Data processing occurs within a multilayered approach, with the health worker retaining control and validating the processed data for a specific period. Following validation, a transaction is recorded on an open ledger, ensuring the immutability of the validated data. The output consists of multi-level reports and real-time monitoring tools. Feedback loops continuously enhance data processing through machine intelligence.

\begin{figure}[H]
    \centering
    \begin{tikzpicture}[scale = 0.95, node distance={15mm}, thick, box/.style = {rectangle, draw}, minimum size = 1cm]



\node[box,color = white, text = black, rotate = 90] at (-10.5,5){\footnotesize Stakeholders};
\node[box,color = white, text = black,rotate = 90] at (-10.5,0.75){Process};
\node[box,color = white, text = black,rotate = 90] at (-10.5,-3.25){\footnotesize Technology};
    \draw[draw=lightgray, rounded corners = 15pt, fill = lightgray] (-10,4) rectangle ++(16,2);

    \node[ellipse,draw, color = lightgray,fill = white, text = black] (a1) at (-5.75cm,4.7cm) {health worker};
    \node[ellipse,draw, color = lightgray,fill = white, text = black] (a2) at (-0.25cm,5cm) {employer};
    \node[ellipse,draw, color = lightgray,fill = white, text = black] (a3) at (2cm,5cm) {\shortstack{approved\\third-party}};

    \draw[draw = lightgray, rounded corners = 15pt, fill = lightgray] (-10,-0.5) rectangle ++(16,2.5);

    \node[box,draw, rounded corners = 15pt, inner sep=7pt,fill = white] (b1) at (-8.5,0.75){\shortstack{multimodal\\input}};

    \node[box,draw,fill = white, inner sep = 10pt] (b2) at (-5.75,0.75){\shortstack{multilayered\\approach}};

    \node[diamond,draw,fill = white] (b3) at (-2.65,0.75) {validation};
    \node[diamond,draw,dotted,fill = white] (b4) at (-0.25,0.75) {checking};
    \node[diamond,draw,dotted,fill = white] (b5) at (2,0.75) {arbitrage};

    \node[box,draw, rounded corners = 15pt, inner sep=12pt,fill = white] (b6) at (4.55,0.75){\shortstack{multilevel\\output}};

    \draw[draw = lightgray, rounded corners = 15pt, fill = lightgray] (-10,-4.25) rectangle ++(16,2);

    \node[box,draw, color = lightgray,fill = white, text = black, inner sep = 15pt] (c1) at (-5.75,-3.25){machine learning};
    \node[box,draw, color = lightgray,fill = white, text = black, inner sep = 15pt] (c2) at (2,-3.25){blockchain ledger};


    \draw[->] (a1) [] to (b2);
    
    \draw[->] (a1) [out = 180, in = 90] to (b1);

    \draw[->] (a1) [out = 0, in = 90] to (b3);

    \draw[->] (b1) to node[midway, yshift = -0.5cm, xshift = 0cm] {
    }(b2);

    \draw[->] (b2) to (b3) to (b4) to (b5) to (b6){};

    \draw[-> , dotted] (a3) [out = -90, in = 90] to (b5){};

    \draw[-> , dotted] (a2) [out = -90, in = 90] to (b4){};

    \node[circle, draw,minimum size = 1pt, color = lightgray] (gg) at (-4.5,5.6){};
    \draw[-, dotted] (a2) [out = 180, in=0, looseness = 1 ] to (gg){};
    \draw[-, dotted] (gg) [out = 180, in=90, looseness = 1 ] to (-8.5,3.55){};
    \draw[-,dotted] (-4.3,5.6) to (-4.6,5.6){};
    \draw[-, dotted] (a3) [out = 110, in=0, looseness = 0.15 ] to (gg){};

\draw[->] (c1) [out = 90, in=-90, looseness = 0.5 ] to (b2);

\draw[->] (c1) [out = 180, in=-90, looseness = 0.5 ] to (b1);

\draw[->] (b3) [out = -90, in=0, looseness = 1 ] to (c1);
\draw[->] (b5) [out = -90, in=0, looseness = 0.5 ] to (c1);
\draw[->] (b3) [out = -90, in=180, looseness = 1 ] to (c2);

\draw[->](b6)  to (4.55,4);
\draw[->](b6)  to (4.55,-2.25);

    \end{tikzpicture} 

\caption{The conceptual framework}
\end{figure}
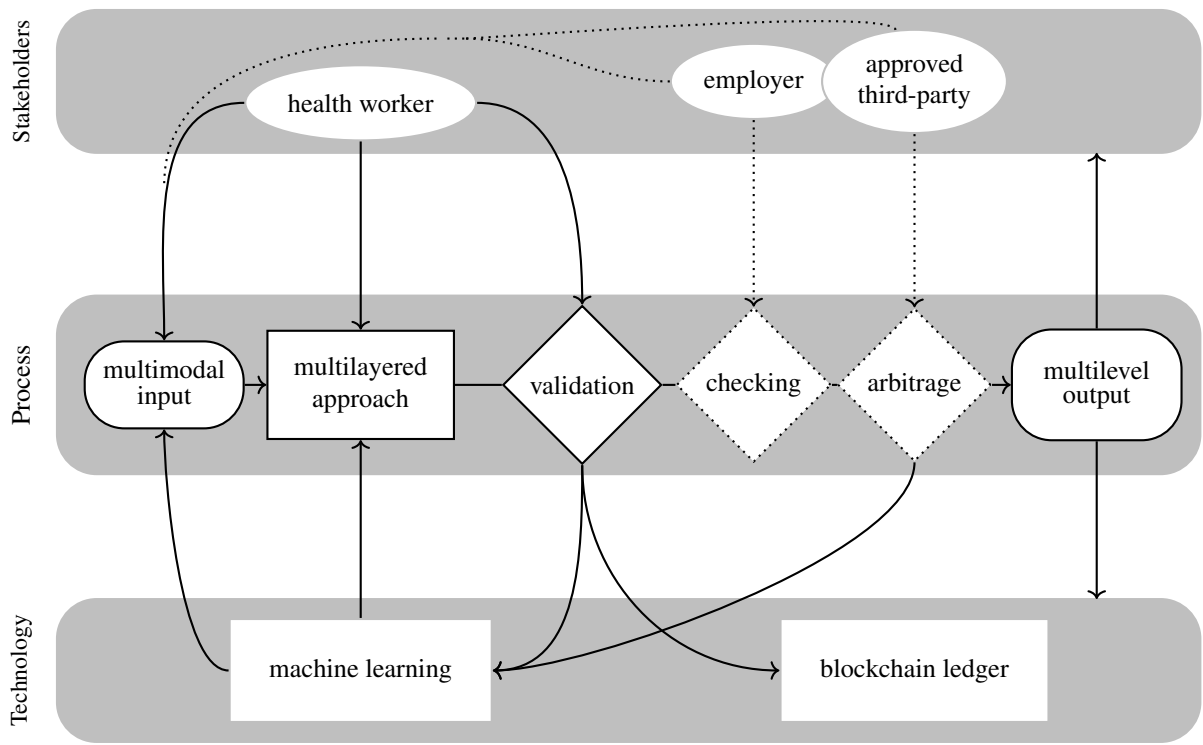

\subsection{Input: multimodal data}

The multimodal data input encompasses a combination of fixed attributes set for predetermined periods (e.g., gross hourly wage) and dynamic information. This dynamic information consists of individual tracking data with collective intelligence insights captured continuously. Timing inputs can be entered synchronously or asynchronously relative to the tracked event, either as a schedule or record.

\begin{table}[H]
\centering
\caption{Examples of timing input modalities with their features.}
\vspace{0.5cm}
\renewcommand{\arraystretch}{3}
\setlength{\tabcolsep}{11.5pt}
\begin{threeparttable} 
\begin{tabular}{c| *5{c|}}
    \cline{2-6} 
    & \textbf{provider}
    & \textbf{timing}
    & \textbf{user friction}
    & \textbf{resilience\tnote{*}} 
    & \textbf{\makecell{infrastructure\\requirement}} \\
    \hline
    \multicolumn{1}{|c|}{\textbf{\makecell{manual\\completion}}} 
    & \makecell{worker or\\employer}
    & async
    & high
    & low
    & low \\
    \hline
    \multicolumn{1}{|c|}{\textbf{\makecell{physical\\badging}}} 
    & worker
    & sync
    & mid
    & mid to high
    & mid to high \\
    \hline
    \multicolumn{1}{|c|}{\textbf{digital log}} 
    & worker
    & sync
    & low to mid
    & mid
    & low to mid \\
    \hline
    \multicolumn{1}{|c|}{\textbf{\makecell{geofencing\\smart contact tracing}}} 
    & process
    & sync
    & low
    & high
    & low to mid \\
    \hline 
\end{tabular}
\begin{tablenotes} 
    \item[*]resilience as the potential ability to withstand internal and external influences.
\end{tablenotes}
\end{threeparttable} 
\end{table}

Timing inputs fall into one of several predefined mutually exclusive states (e.g., "at work," "available," or "on leave"). Indicators can be used to label or tag these inputs. Labels are predefined standardized sub-options tailored for collective needs (e.g., "care," "education," "research," and "administrative" for the "at work" state; "on-call" and "second-line" for the "available" state; "annual leave," "scientific leave," or "sickness leave" for the "on leave" state). Tags are free-form keywords defined by individual users for their specific needs. Additionally, indicators include sporadic event tracking, which can be either user-driven (e.g., incident reporting) or solution-driven (e.g., mood monitoring).

\subsection{Process: a multi-layered approach}

Input integration employs a tiered approach that combines various layers of timing data, such as schedules, time tracking, and records, along with complementary indicators. These layers serve as safety nets for missing information and can be tailored or deemed obligatory in accordance with specific needs or legal requirements.

\begin{figure}[H]
\centering
\begin{tikzpicture}[scale=.9,every node/.style={minimum size=1cm},on grid, point/.style = {circle,fill=black,inner sep=0pt,minimum size=3pt}]
		

\draw[>-Stealth, line width = 1mm](0,-7.78)to(0,13.03){};

    \begin{scope}[
    	yshift=-200,every node/.append style={
    	    yslant=0.5,xslant=-1},yslant=0.5,xslant=-1
    	             ]
        \foreach \i [count=\n] in {0,1,...,20}{ 
            \pgfmathsetmacro{\x}{5*int(random(0,6))}
            
            \pgfmathsetmacro{\c}{int(random(0,256)}
            \pgfmathsetmacro{\v}{int(random(1,4)*5}
            \pgfmathsetmacro{\y}{5*int(random(0,9.5))}
            \fill[lightgray!\c] (\x mm,\y mm) rectangle ++ (\v mm,5 mm);
        }
        \draw[step=5mm, white, line width=3pt] (0,0) grid (5,5);
		\draw[step=5mm, black] (0,0) grid (5,5);
		\draw[white,very thick] (0,0) rectangle (5,5);
        \draw[black,very thick,dashed] (0,0) rectangle (5,5);
    \end{scope}

    \begin{scope}[yshift=-100,every node/.append style={yslant=0.5,xslant=-1},yslant=0.5,xslant=-1]
        \fill[white,fill opacity=.5] (0,0) rectangle (5,5);
         \foreach \i [count=\n] in {0,1,...,20}{ 
            \pgfmathsetmacro{\x}{5*int(random(0,6))}
            
            \pgfmathsetmacro{\c}{int(random(0,256)}
            \pgfmathsetmacro{\v}{int(random(1,4)*5}
            \pgfmathsetmacro{\y}{5*int(random(0,9.5))}
            \fill[lightgray!\c] (\x  mm,\y  mm) rectangle ++ (\v mm, 5 mm);
        }
        \draw[step=5mm, white, line width=3pt] (0,0) grid (5,5);
        \draw[step=5mm, black] (0,0) grid (5,5);
        \draw[white,very thick] (0,0) rectangle (5,5);
        \draw[black,dashed] (0,0) rectangle (5,5);
  
    \end{scope}
    \begin{scope}[yshift=0,every node/.append style={yslant=0.5,xslant=-1},yslant=0.5,xslant=-1 ]
         \foreach \i [count=\n] in {0,1,...,20}{ 
            \pgfmathsetmacro{\x}{5*int(random(1,5))}
            
            \pgfmathsetmacro{\c}{int(random(0,256)}
            \pgfmathsetmacro{\v}{int(random(1,4)*5}
            \pgfmathsetmacro{\y}{5*int(random(1,8.5))}
            \fill[lightgray!\c] (\x mm,\y mm) rectangle ++ (\v mm,5 mm);
        }
        \draw[step=5mm, white, line width=3pt] (0.5,0.5) grid (5,5);
        \draw[step=5mm, black] (0,0) grid (5,5);
        \draw[-,color = white, line width= 1pt](0,0)to (5,0){};
        \draw[-,color = white, line width= 1pt](5,0)to (5,5){};
        \draw[-,color = white, line width= 1pt](5,5)to (0,5){};
        \draw[-,color = white, line width= 1pt](0,0)to (0,5){};

    \end{scope}
    \begin{scope}[yshift=100,every node/.append style={yslant=0.5,xslant=-1},yslant=0.5,xslant=-1]
        \fill[white,fill opacity=0.5] (0.1,0.1) rectangle (5,5);
           \foreach \x in {0.2,0.4,...,4.8}
           \foreach \y in {0.2,0.4,...,4.8}
           \fill[gray] (\x,\y) circle[radius=1.5pt];
        \draw[step=1mm, darkgray,thin] (3,1) grid (4,2);  

        \fill[darkgray] (3,1) circle (0.08);
        \fill[darkgray] (3,1.2) circle (0.08);
        \fill[darkgray] (3.2,1.2) circle (0.08);
        \fill[darkgray] (3.6,1.2) circle (0.08);
        \fill[darkgray] (3.6,1.6) circle (0.08);
        \fill[darkgray] (3.6,2) circle (0.08);
        \fill[darkgray] (3.6,2.2) circle (0.08);
        \fill[darkgray] (3.2,2) circle (0.08);
        \fill[darkgray] (4,2) circle (0.08);
        \fill[darkgray] (4.4,1.4) circle (0.08);
        \draw [->][black] (4.4,1.4) -- (4,2);
        \draw [->][black] (4.4,1.4) -- (3.6,2);
        \draw [->][black] (4.4,1.4) -- (3.2,2);
        \draw [->][black] (4.4,1.4) -- (3,1);
        \draw [->][black] (3,1) -- (3.6,2.2);
        \draw [->][black] (3.6,1.6) -- (3.6,1.2);
        \draw [->][black] (3.6,1.6) -- (3,1.2);
        \draw[step=1mm, black,thin] (4,4) grid (4.5,4.55);  
        \fill[black] (4.4,4.4) circle (0.08);
        \fill[black] (4.2,4.2) circle (0.08);
        \draw [->][black] (4.4,4.4) -- (3.6,2.2);

        \draw[step=1mm, darkgray,thin] (1,3) grid (2,4);  
        \draw [->][black] (1.4,4.4) -- (2,4);
        \draw [->][black] (1.4,4.4) -- (2,3.6);
        \draw [->][black] (1.4,4.4) -- (2,3.2);
        \draw [->][black] (1.4,4.4) -- (1,3);
        \draw [->][black] (1,3) -- (2.2,3.6);
        \draw [->][black] (1.6,3.6) -- (1.2,3.6);
        \draw [->][black] (1.6,3.6) -- (1.2,3);
        \fill[black] (1,3) circle (0.08);
        \fill[black] (1.2,3) circle (0.08);
        \fill[black] (1.2,3.2) circle (0.08);
        \fill[black] (1.2,3.6) circle (0.08);
        \fill[black] (1.6,3.6) circle (0.08);
        \fill[black] (2,3.6) circle (0.08);
        \fill[black] (2.2,3.6) circle (0.08);
        \fill[black] (2,3.2) circle (0.08);
        \fill[black] (2,4) circle (0.08);
        \fill[black] (1.4,4.4) circle (0.08);
        
    \end{scope}

 \begin{scope}[
    	yshift=200,every node/.append style={
    	    yslant=0.5,xslant=-1},yslant=0.5,xslant=-1
    	             ]
        \fill[white,fill opacity=.5] (0,0) rectangle (5,5);
        \foreach \i [count=\n] in {0,1,...,20}{ 
            \pgfmathsetmacro{\x}{5*int(random(0,6))}
            
            \pgfmathsetmacro{\c}{int(random(0,256)}
            \pgfmathsetmacro{\v}{int(random(1,4)*5}
            \pgfmathsetmacro{\y}{5*int(random(0,9.5))}
            \fill[lightgray!\c] (\x mm,\y mm) rectangle ++ (\v mm,5 mm);
        }
        \draw[step=5mm, black] (0,0) grid (5,5);
        \draw[black,very thick] (0,0) rectangle (5,5);

        \node[point](1) at (1.25,0.75) {};
        \node[point](2) at (2.25,0.75) {};
        \node[point](3) at (2.75,0.75) {};
        \node[point](4) at (2.25,1.75) {};
        \draw[-] (1) to (2) to (3);
        \draw[-] (2) to (4);
        \node[point](5) at (3.25,1.75) {};
        \node[point](6) at (3.75,1.25) {};
        \draw[-] (5) to (6);
        \node[point](7) at (1.25,2.75) {};
        \node[point](8) at (1.25,3.25) {};
        \node[point](9) at (2.25,3.25) {};
        \node[point](10) at (2.25,4.25) {};
        \draw[-] (7) to (8) to (9) to (10);

    \end{scope}
\draw[-Stealth, line width = 1mm](0,11.75)to(0,13.03){};
    
    \draw[-latex,thick](-4.5,7)node[left]{$\mathsf{indicators}$}
        to[out=0,in=90] (-2,6.5);

    \draw[-,thick](6,4.5)node[right]{$\mathsf{machine \ intelligence}$}
        to[out=180,in=90] (3.6,5.5);

    \draw[-,thick](6,-2.5)node[right]{$\mathsf{provisional \ schedule}$}
        to[out=180,in=90] (4,-1.5);

    \draw[-,thick](6,1)node[right]{$\mathsf{time \ tracking}$}
        to[out=180,in=90] (4,2);
    
    \draw[-,thick](6,-6)node[right]{$\mathsf{default \ schedule}$}
        to[out=180,in=90] (4,-5);
     
     \draw[-,thick](6,8)node[right]{$\mathsf{validated \ record}$}
     to[out=180,in=90] (4,9);
     
     \draw[->,thick](-4.5,-5.9)node[left]{$\mathsf{state}$}
     to[out=0,in=90] (-3,-5.3);

\end{tikzpicture}

\caption{The multilayered approach for multimodal data integration enhanced by machine intelligence.}
\end{figure}
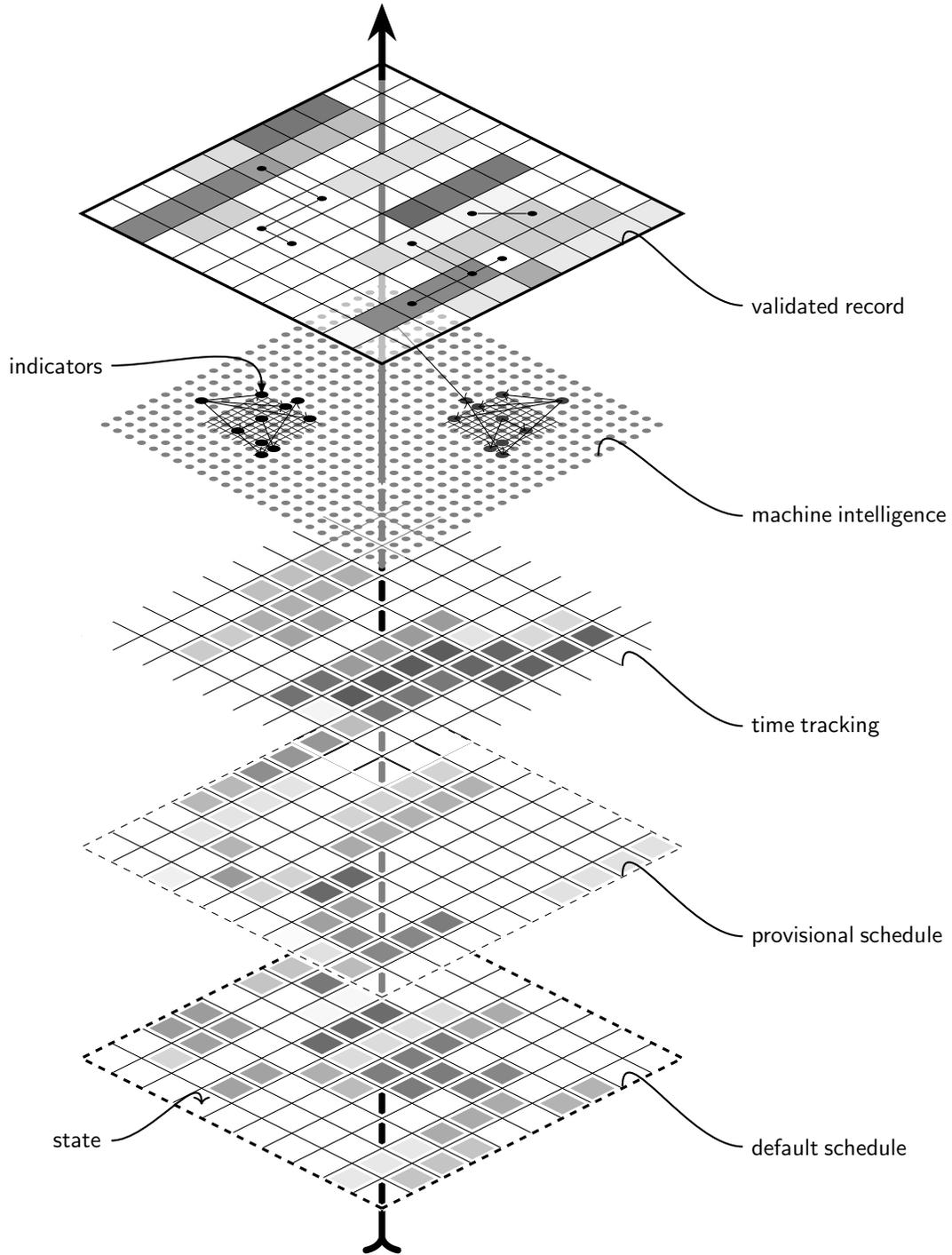

To address conflicting information between these layers, the solution offers four automated resolution options: union, intersection, superseding layer, or a machine learning (ML) forecast.

\begin{figure}[H]
\centering
    \begin{tikzpicture}[scale = 0.8]
        \draw[->] (-7,3.5) to (6.5,3.5){};
        \draw[-] (-12.5,0) to (-8,0){};
        \draw[-,color = black](-7,0) to (6.5,0){};
        
        \foreach \i [count=\n] in {6,7,...,17}{
        \draw[-] (\i-13,3.4) to (\i -13,3.6){}; 
        \node[font=\scriptsize, text height=1.75ex,text depth=.5ex] at (\i-13,3.8) {$\i$};
        }
        \node[font = \scriptsize, text height = 1.75ex, text depth = .5ex] at (5.5,3.8){day time};
        \node[] at (-10.93,2.75){default schedule};
        \draw[pattern={vertical lines}] (-5,2.5) rectangle ++(7,0.5);

         \node[] at (-10.59,1.75) {provisional schedule};
        \draw[pattern={north east lines}] (-4,1.5) rectangle ++(5,0.5);
        
        \node[] at (-11.21,0.75){time tracking};
        \draw[pattern={north west lines}] (-6.5,.5) rectangle ++(1,0.5);
        \draw[pattern={north west lines}] (-4,.5) rectangle ++(3,0.5);
        \draw[pattern={north west lines}] (0,.5) rectangle ++(4.5,0.5);
        
        \node[] at (-10.40,-0.75){union without merging};
        \draw[pattern={north west lines}] (-6.5,-1) rectangle ++(1,0.5);
        \draw[pattern={north west lines}] (-4,-1) rectangle ++(3,0.5);
        \draw[pattern={north west lines}] (0,-1) rectangle ++(4.5,0.5);
        \draw[pattern={north east lines}] (-4,-1) rectangle ++(5,0.5);
        \draw[pattern={vertical lines}] (-5,-1) rectangle ++(7,0.5);
        
        \node[] at (-10.7,-1.75){union with merging};
        \draw[pattern={north west lines}] (-6.5,-2) rectangle ++(1,0.5);
        \draw[pattern={north west lines}] (-4,-2) rectangle ++(3,0.5);
        \draw[pattern={north west lines}] (0,-2) rectangle ++(4.5,0.5);
        \draw[pattern={north east lines}] (-4,-2) rectangle ++(5,0.5);
        \draw[pattern={vertical lines}] (-5,-2) rectangle ++(7,0.5);
        \draw[](-5.5,-2) rectangle node[align = center]{$\ast$}++ (0.5,0.5) ;
        
        \node[] at (-11.37,-2.75){intersection};
        \draw[pattern={north west lines}] (-4,-3) rectangle ++(3,0.5);
        \draw[pattern={north west lines}] (0,-3) rectangle ++(1,0.5);
        \draw[pattern={north east lines}] (-4,-3) rectangle ++(3,0.5);
        \draw[pattern={north east lines}] (0,-3) rectangle ++(1,0.5);
        \draw[pattern={vertical lines}] (-4,-3) rectangle ++(3,0.5);
        \draw[pattern={vertical lines}] (0,-3) rectangle ++(1,0.5);

        \node[] at (-11.315,-3.75){ML forecast};
\draw[pattern={north west lines}] (-6.5,-4) rectangle ++(1,0.5);
\draw[pattern={north west lines}] (-4,-4) rectangle ++(3,0.5);
\draw[pattern={north west lines}] (0,-4) rectangle ++(4.5,0.5);
\draw[pattern={north east lines}] (-4,-4) rectangle ++(3,0.5);
\draw[pattern={north east lines}] (0,-4) rectangle ++(1,0.5);
\draw[pattern={vertical lines}] (-5,-4) rectangle ++(4,0.5);
\draw[pattern={vertical lines}] (0,-4) rectangle ++(2,0.5);
\draw[](-5.5,-4) rectangle node[align = center]{$\ast$} ++(0.5,0.5);

\draw(-1,-4) -- (-1,-3.5);
\draw(0,-4) -- (0,-3.5);

\draw(-1,-3.5) -- (-1,-3.5);
\draw(-0,-3.5) -- (0,-3.5);
\draw(-1,-4) -- (-1,-4);
\draw(-0,-4) -- (0,-4);

\node at (-0.5, -3.75) {\textreferencemark};

    \end{tikzpicture}
\smallskip
\smallskip
\smallskip
\\
\footnotesize $\ast$ recovered missing working period; \textreferencemark{} 
 detected lunch pause.
\smallskip
\smallskip
\smallskip
\caption{Examples of timing conflicts resolution approaches}
\label{fig:timing_conflicts}
\end{figure}
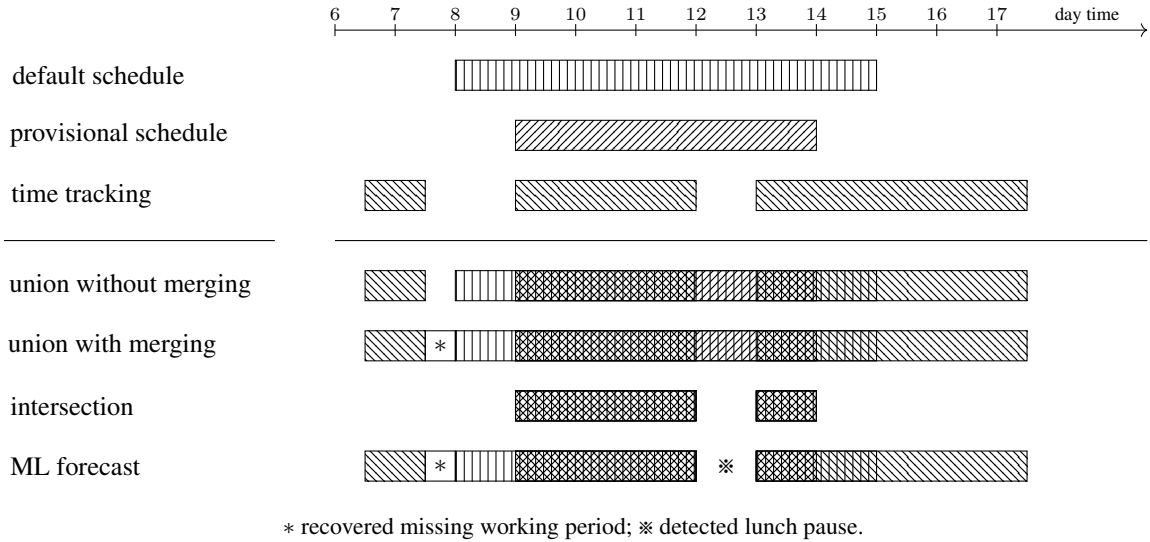

Health workers have the ability to review and validate their time recordings on a punctual basis. Optionally, they can place the validated recording on hold for an approval period, allowing employers or authorized third parties to review it. In case of an unresolved dispute, the pending litigations are described in the final report. This report is then securely and immutably recorded on the blockchain.

\subsection{Output: multi-level reporting}

At the individual level, a dashboard displays time tracking, wage estimates, occupational safety, health monitoring, and legal compliance metrics. Users can generate on-demand static reports to summarize validated data for a specified period with automated email reports. At the collective level, the solution provides overview reports based on geographic or organizational criteria, such as institutional, local, or global levels. These collective reports are made openly accessible on a dedicated platform for all stakeholders and the general public.

\begin{table}[H]
\centering
\caption{Examples of insights,  metrics, indices, and alerts of interest.}
\vspace{0.5cm}

\renewcommand{\arraystretch}{1.5}
\begin{tabular}{|l|l|l|l|}
\hline
\textbf{Work}         & \textbf{Safety}  & \textbf{Wellbeing}     & \textbf{Wage}      \\ \hline
working time          & shift durations  & mood tracker  & salary calculator \\ \hline
overtime              & resting time     & burnout index & extra pay         \\ \hline
unsociable hours      & legal compliance & job satisfaction        & salary comparisons    \\ \hline
activity tracking     & commute risk         & incident reporting             &  professional expenses                 \\ \hline
\end{tabular}
\end{table}

\section{Governance}

A decentralized autonomous organization (DAO) \cite{EthereumSite}  \cite{Weyl2022} \cite{jentzsch2016decentralized} facilitates distributed governance, enabling all stakeholders to participate equitably in decision-making and oversight of operations. Governance tokens act as a voting mechanism, featuring dynamic supply adjustments to prevent single-entity dominance \cite{Plutus}. Token distribution considers the varying sizes and influences of participants and allows individuals to delegate their voting power to trusted organizations. Sybil attacks – i.e., fraudulent creation of multiple accounts for increased voting power – are prevented by identification and authentication processes for user verification with rewarding protocols relying on proof of attendance, effort, and contribution \cite{Sizon2018}. SubDAOs, functioning as independent DAOs within the larger organization, cater to specific endeavors. Developers can collaborate to address challenges such as feature development, technological enhancement, solution maintenance, and security monitoring. Besides, local stakeholders from specific regions or institutions can organize daily operations, dispute resolution, and policymaking, as well as request customizations to meet their unique requirements. This structure complements the delegation system. 

\begin{figure}[H]
	\centering
	\begin{tikzpicture}
		\path[mindmap,concept color = lightgray!50!white,text=white, grow cyclic,
		level 1/.append style = {
			level distance = 4.5cm,
			sibling angle = 90
		},
		level 2/.append style = {
			level distance = 3cm,
			sibling angle = 45},
		level 3/.append style = {
			level distance = 2cm,
			sibling angle = 45}]
		node[concept, text = black] {DAO} [clockwise from= - 45]
		child[concept color = gray!50!lightgray] {
			node[concept] {developer subDAO}[clockwise from = 45]
			child[concept color = darkgray!50!gray] { 
				node[concept] {maintenance} }
			child[concept color = darkgray!50!gray] { 
				node[concept] {innovation} }
			child[concept color = darkgray!50!gray] { 
				node[concept] {security} }
		} 
		child[concept color = gray!50!lightgray] { 
			node[concept] {local subDAO}[clockwise from = -135]
			child[concept color = darkgray!50!gray]{ 
				node[concept,text= white] {deployment}}
			child[concept color = darkgray!50!gray]{ 
				node[concept, text = white] {governance}[clockwise from = 145]}
			child[concept color = darkgray!50!gray]{ 
					node[concept]{arbitrage}}
};
\end{tikzpicture}

\caption{Example of a DAO-subDAO structure for synergistic yet independent working groups.}
\end{figure}
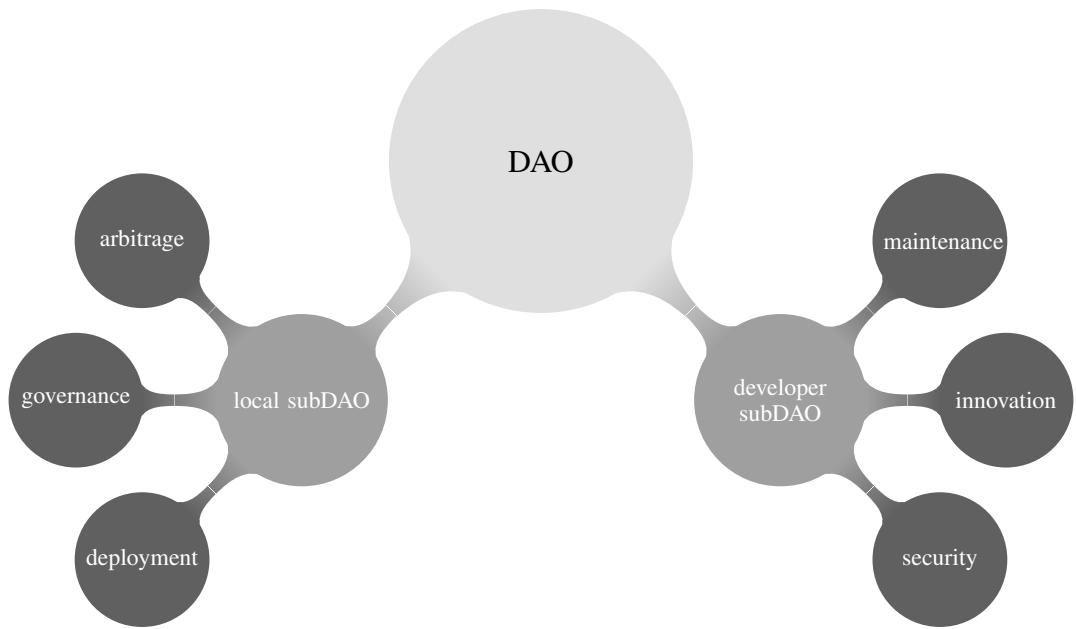

\section{Legal frameworks}
The solution ensures compliance with governmental working time regulations and specific recommendations by generating derived metrics for informing health workers and stakeholders. In terms of data protection, it upholds privacy and security best practices. Key principles, such as purpose and storage limitations, are observed by processing data solely for its intended use and retaining it only for the necessary duration. Additionally, data minimization, pseudonymization, and encryption techniques are employed. Sharing aggregated personal data is contingent upon obtaining explicit user consent. Secure machine intelligence protocols, including federated learning, are incorporated to guarantee user privacy and data security, addressing related legal concerns.

\section{Technological stack}

The stack utilizes accessible open-source tools in a decentralized framework, promoting transparency and collaborative development. Crowdsourcing approaches engage various perspectives and innovations, bolstering the system's flexibility, adaptability, and versatility. Machine learning further empowers the solution to improve its capabilities continually in response to evolving user needs.

\subsection{Versatile architecture}

This robust, distributed, and privacy-preserving technological stack integrates multi-factor authentication, end-to-end encryption, blockchain technology, and machine intelligence. Users authenticate their identities using mobile phone-based two-factor authentication via a decentralized authentication protocol \cite{Yasuda2022} with electronic identification and signature. Data transmission security is ensured through end-to-end encryption, safeguarding sensitive information by allowing access exclusively to intended recipients. Blockchain technology offers a decentralized, transparent platform for data storage and sharing, guaranteeing data immutability, trusted timestamping, and tamper resistance. Furthermore, state-of-the-art blockchains feature Turing-complete smart contract capabilities, extensive decentralized applications ecosystems, and native interoperability. These blockchains provide virtual machines, enabling the execution of complex, logic-based smart contracts and applications designed to address the wide range of our use cases and requirements \cite{EthereumSite} \cite{polkadot2016}.

\subsection{Machine intelligence}

Recurrent neural networks with long short-term memory units effectively analyze sequential data patterns, such as timing input resolution, work schedule optimization but also burnout prevention, and mood monitoring through sentiment analysis \cite{Long1997}. Anomaly detection identifies potential issues like overwork, resource misuse, and legal compliance violations \cite{Chandola2009}. Additionally, convolutional neural networks complement these applications by managing image recognition and processing tasks involving multidimensional data, such as user interaction patterns, behavioral signals, and image-based input processing \cite{Yamashita2018}. Adhering to the decentralized approach and privacy concerns, federated learning provides a secure and privacy-preserving method for analyzing aggregated user data. This technique enables the processing of individual data locally on the user's device or through anonymization and encryption before transmission, allowing global models to be trained without sharing identifiable raw data. Customized deep-learning models can be developed using available open-source frameworks suited to the solution's requirements \cite{tensorflow2015-whitepaper} \cite{DBLP:journals/corr/abs-1811-04017} \cite{DBLP:journals/corr/abs-1902-01046} \cite{DBLP:journals/corr/abs-1912-01703}.

\section{Implementation}

The solution’s rollout requires addressing technical deployment challenges while promoting early adoption among users and stakeholders. To this end, we aim to capitalize on its agile infrastructure and innovative approach, tackling an unaddressed but widespread and longstanding issue in healthcare.

\subsection{Technical deployment}

Scalability is essential for accommodating an increasing user base and extending features. A modular architecture, employing microservices that integrate cloud and on-premises open-source software, can facilitate efficient scaling and performance while meeting user-specific needs and requirements. Interoperability, through standardized data exchange protocols, enables seamless integration with existing health information systems and time-tracking platforms. We propose a simplified time-tracking file in CSV format as a standard. 

\begin{table}[H]
\centering
\caption{Examples of standardized fields and values for activity tracking}
\vspace{0.5cm}
\renewcommand{\arraystretch}{1.5}
\begin{tabular}{|p{6cm}|p{6cm}|}
\hline
\textbf{Fields}        & \textbf{Description}     \\ \hline
location    &  latitude and longitude as decimal numbers             \\ \hline
employer    &  unique ID of employer          \\ \hline
start/end\_date    & the date when the tracked activity begins/ends, in the format YYYY-MM-DD    \\ \hline
start/end\_time    & the time when the tracked activity begins/ends, in the format HH:MM:SS \\ \hline
state/label/tag   & according to customized lists                   \\ \hline
\end{tabular}
\end{table}

Cloud-based processes and cross-platform compatibility further enhance the solution's versatility. Customization, enabled by adaptable application programming interfaces (APIs), permits tailoring the solution to particular needs. Security must be maintained by thorough testing, periodic audits, and ongoing monitoring to protect sensitive data and ensure privacy regulation compliance. A dedicated subDAO will oversee these endeavors supported by the open-source community.

\subsection{Early adoption}

This solution bridges a technological gap in healthcare, addressing a critical unmet need for health workers by offering a versatile platform for individual and collective insights into occupational management. To promote early adoption, we will target individual users, particularly showcasing the platform's features and benefits for individuals. Creating momentum, we aim to incentivize diverse stakeholders to onboard subsequently, fully leveraging the potential of the platform. Moreover, open beta testing with pilot programs will provide real-world validation and insights for refining the solution, demonstrating its viability, and encouraging further adoption and retention. Engaging stakeholders, such as health workers' associations, governmental organizations, and healthcare employers, is essential for broad acceptance and impact. Along these lines, distributed and transparent governance aims to align often conflicting interests by maintaining the priority on improving health workers' lives and working conditions. Stakeholder engagement through workshops and seminars will foster shared ownership and address diverse perspectives. Coupled with incentivization via gamification and reward systems, this approach will motivate active engagement, contributing to long-term success. Additionally, providing training and support will empower users to effectively navigate the solution, ensuring maximum utility and benefits for all parties involved.

\section{Conclusion}
We have introduced a comprehensive worker-centered solution to support health workers facing pressing issues of safety, well-being, and productivity. By harnessing machine intelligence, this solution integrates multimodal data through a multilayered approach, enhancing user experience. Blockchain technology enables distributed governance and transparency, fostering effective cooperation among stakeholders. We have demonstrated the technology's readiness for deployment in compliance with the most stringent openness, decentralization, and privacy-preserving standards. Overcoming implementation hurdles is critical to realizing the solution's full potential and delivering a lasting, transformative impact on public health. This solution can bridge an unaddressed technological gap in healthcare, offering a unique opportunity to incentivize health workers' engagement and stakeholders' alignment in developing occupational management for better care. 

\section*{Acknowledgements}
We express our gratitude to Maaz and Syed Ali Shahbaz from \href{https://www.timecrypt.app}{Timecrypt} for our collaboration on this project. Additionally, we extend our thanks to the Belgian junior doctors from \href{https://www.ladelegation.be}{la Délégation des Médecins Francophones en Formation} for their diligent alpha testing of the prototype app and their critical insights.


\printbibliography
\end{document}